*Correspondence and requests for materials should be addressed to M.Q. (email: qinmh@scnu.edu.cn) or J.-M.L. (email: liujm@nju.edu.cn)

# Helical and skyrmion lattice phases in three-dimensional chiral magnets: Effect of anisotropic interactions

J. Chen[1], W. P. Cai[1], M. H. Qin[1,*], S. Dong[2], X. B. Lu[1], X. S. Gao[1], and J. -M. Liu[3]

[1]Institute for Advanced Materials, South China Academy of Advanced Optoelectronics and Guangdong Provincial Key Laboratory of Quantum Engineering and Quantum Materials, South China Normal University, Guangzhou 510006, China

[2]Department of Physics, Southeast University, Nanjing 211189, China

[3]Laboratory of Solid State Microstructures, Nanjing University, Nanjing 210093, China

**In this work, we study the magnetic orders of the classical spin model with the anisotropic exchange and Dzyaloshinskii-Moriya interactions in order to understand the uniaxial stress effect in chiral magnets such as MnSi. Variational zero temperature ($T$) calculated results demonstrate that various helical orders can be developed depending on the magnitude of the interaction anisotropy, consistent with the experimental observations at low $T$. Furthermore, the creation and annihilation of the skyrmions by the uniaxial pressure can be also qualitatively reproduced in our Monte Carlo simulations. Thus, our work suggests that the interaction anisotropy tuned by applied uniaxial stress may play an essential role in modulating the magnetic orders in strained chiral magnets.**

In the past a few years, the nontrivial magnetic orders observed in chiral magnets such as MnSi [1-3], $Fe_{1-x}Co_xSi$ [4] and FeGe [5, 6] have been attracting continuous attentions due to their interesting physics and potential applications for future memory technology. Specifically, a helical order with a single ordering wave vector **k** (point along the [111] axis in MnSi, for example) is developed at low temperatures (*T*) under zero magnetic field (*h*), resulting from the competition between the ferromagnetic (FM) exchange interaction and the Dzyaloshinskii-Moriya (DM) interaction [7, 8]. When a finite *h* is applied, the helical order is replaced by the conical phase to save the Zeeman energy. More interestingly, a skyrmion lattice phase [9] (with a vortex-like spin configuration where the spins point in all directions forming a sphere) is stabilized in a certain (*T*, *h*) region, and is proposed to be potentially used for data encoding because of its efficient modulation by ultralow current density [10, 11] (~$10^5$ – $10^6$ A $m^{-2}$, orders of magnitude smaller than that for domain-wall manipulation) and its topological stability. Theoretically, the cooperation of the energy competition (among the FM, DM, and Zeeman couplings) and thermal fluctuations is suggested to contribute to the stabilization of the skyrmion lattice phase [12] in bulk chiral magnets, and Rashba spin-orbit coupling in two-dimensional materials is believed to further enhance the stability of skyrmions [13].

Subsequently, a number of theoretical simulations searching for effective manipulation methods of skyrmions have been performed in order to develop related spintronic devices. It is suggested that skyrmions in bulk and/or thin film systems could be controlled by external stimuli such as electric currents [14], magnetic fields [15] and thermal gradients [16, 17]. As a matter of fact, some of these manipulations have been confirmed in experiments [18], although it is very hard to create and annihilate skyrmions using these methods [19].

Most recently, the dependence of the magnetic orders on uniaxial pressure in MnSi has been investigated experimentally in detail [20, 21]. The wave vector of the helical order at zero *h* is reoriented from the [111] axis to the stress axis when the uniaxial pressure is applied. More importantly, the *T*-region of the skyrmion lattice phase can be extensively modulated by the pressure, demonstrating a new manipulation method in this system. Specifically, the extent of the skyrmion lattice phase is strongly enhanced for pressures applied perpendicular to the magnetic field, while is slowly decreased under pressures parallel to the field. So far, the microscopic mechanism for the strain effect is not clear, and urgently deserves to be uncovered in order to understand the physics and even speed up the application process [22].

Fortunately, the earlier spin model has successfully reproduced the ordered phases found in the experiments on bulk MnSi, allowing one to explore the strain effect based on a modified model [12]. Usually, uniaxial pressures may enhance lattice distortion, and in turn modulate exchange anisotropies in a magnetic system [23, 24]. For example, exchange anisotropy has been proven to be very important in the strained manganite thin films [25, 26] and in strained iron pnictides [27, 28]. Furthermore, the DM interaction in chiral magnets along the compressive axis is found to be largely enhanced when a pressure is applied, as revealed in earlier experiments (on FeGe thin films) [24] and first-principles calculations (on $Mn_{1-x}Fe_xGe$) [29]. Thus, it is essential to make clear the role of the interaction anisotropy in modulating the magnetic orders in order to understand the strain effect in chiral magnets. More importantly, such study may provide useful information about the magnetic orders in similar magnets with anisotropic interactions.

In this work, we study a classical Heisenberg spin model including anisotropic FM exchange and DM interactions on a three-dimensional lattice by combining variational zero $T$ calculations with Monte Carlo (MC) simulations to understand the strain effect on magnetic orders in bulk MnSi. The experimentally reported reorientation of the wave vector of the helical order and the variation of the extent of skyrmion lattice phase in experimental phase diagrams under uniaxial stress are qualitatively reproduced when the interaction anisotropies are taken into account.

**Model and methods**

**In** this work, the modified spin model is used to describe strained MnSi, and its Hamiltonian is given by:

$$H = -\sum_{\mathbf{i}} (J_x \mathbf{S_i} \cdot \mathbf{S_{i+\hat{x}}} + J_y \mathbf{S_i} \cdot \mathbf{S_{i+\hat{y}}} + J_z \mathbf{S_i} \cdot \mathbf{S_{i+\hat{z}}})$$
$$- \sum_{\mathbf{i}} (D_x \mathbf{S_i} \times \mathbf{S_{i+\hat{x}}} \cdot \hat{x} + D_y \mathbf{S_i} \times \mathbf{S_{i+\hat{y}}} \cdot \hat{y} + D_z \mathbf{S_i} \times \mathbf{S_{i+\hat{z}}} \cdot \hat{z}) - h \sum_{\mathbf{i}} S_\mathbf{i}^z , \quad (1)$$

where $\mathbf{S_i}$ represents the Heisenberg spin with unit length on site $\mathbf{i}$, $\hat{x}$, $\hat{y}$, $\hat{z}$ are the basis vectors of the cubic lattice. The first term is the anisotropic FM exchange interaction between the nearest neighbors with the interaction constant $J_\mu$ ($\mu = x, y, z$). The second term is the anisotropic DM interaction with $D_\mu$ ($\mu = x, y, z$). The last term is the Zeeman coupling with $h$ applied along the [001] direction. For simplicity, $J_x$, $J_y$, the lattice constant, and the Boltzmann constant are set to unity. In this work, the ground states are obtained with an analytical approach, and the finite-$T$ phase diagrams are estimated by MC simulations. It is noted that the system size studied in this work is much larger than that of the skyrmion, and the demagnetization energies which are important in nanostructures [30, 31] can be safely ignored comparing with the DM and FM couplings [32].

In isotropic bulk system under zero $h$, the ground state is the helical order with the wave vector [8] $\mathbf{k} = \arctan(D/\sqrt{3} J) (1, 1, 1)$ of which the orientation is usually attributed to weak magneto-crystalline anisotropy [33, 34]. Furthermore, uniaxial anisotropy also can efficiently modulate the magnetic states in chiral magnets [35, 36] and other magnetic materials [37, 38], as revealed in earlier works. However, exact solution of the model further considering magneto-crystalline anisotropy is hard to be calculated using the variational method. Thus, such anisotropy is not considered here in order to help one to understand the effect of interaction anisotropy clearly, and our physical conclusions are not affected by this ignorance. Interestingly, when an interaction anisotropy is considered, the ground-state of the system is still a single-$\mathbf{k}$ helical order with $\mathbf{k} = (k_x, k_y, k_z)$, as will be explained latter. Without loss of generality, we set the rotation axis vector $\mathbf{R}$ and initial spin $\mathbf{S_0}$, respectively, to be:

$$\mathbf{R} = (\sin\varphi\cos\theta, \sin\varphi\sin\theta, \cos\varphi)$$
$$\mathbf{S_0} = (\cos\varphi\cos\theta, \cos\varphi\sin\theta, -\sin\varphi) . \quad (2)$$

Then, the spin vector $\mathbf{S_i}$, the energy per site $E$ and effective field $\boldsymbol{f_i}$, respectively, can be calculated by:

$$\mathbf{S_i} = \sin(\mathbf{k} \cdot \mathbf{i}) \cdot \mathbf{R} \times \mathbf{S_0} + \cos(\mathbf{k} \cdot \mathbf{i}) \cdot \mathbf{S_0} \tag{3}$$

$$E = -(J_x \cos k_x + J_y \cos k_y + J_z \cos k_z) \\ - (D_x \sin\varphi\cos\theta \sin k_x + D_y \sin\varphi\sin\theta \sin k_y + D_z \cos\varphi \sin k_z), \tag{4}$$

and

$$\mathbf{f_i} = -\frac{\delta H}{\delta \mathbf{S_i}} = \sum_\mu J_\mu \mathbf{S_{i+\hat{\mu}}} + \sum_\mu D_\mu \mathbf{S_{i+\hat{\mu}}} \times \hat{\mu}. \tag{5}$$

By optimizing for $\mathbf{k}$ and ($\theta$, $\varphi$), we obtain the following equation set:

$$\begin{cases} J_x \tan k_x = D_x \sin\varphi\cos\theta \\ J_y \tan k_y = D_y \sin\varphi\sin\theta \\ J_z \tan k_z = D_z \cos\varphi \\ D_z \sin k_z \tan\varphi = D_x \cos\theta \sin k_x + D_y \sin\theta \sin k_y \\ D_x \sin\varphi\sin\theta \sin k_x = D_y \sin\varphi\cos\theta \sin k_y \end{cases} \tag{6}$$

Here, the last two equations ensure $\mathbf{S_i} \times \mathbf{f_i} = 0$, confirming that the single-$\mathbf{k}$ helical order is the ground state. Then, we can uncover the ground-state of the system at zero $h$ for determined $D_\mu$ and $J_\mu$ through energy analysis.

In addition, the finite-$T$ phase diagram under various $h$ is also calculated by MC simulations. Following the earlier work [12], a compensation term is considered in the model Hamiltonian to minimize the discretization errors in the simulations, which can be given by:

$$H_C = \frac{1}{16} \sum_\mathbf{i} (J_x \mathbf{S_i} \cdot \mathbf{S_{i+2\hat{x}}} + J_y \mathbf{S_i} \cdot \mathbf{S_{i+2\hat{y}}} + J_z \mathbf{S_i} \cdot \mathbf{S_{i+2\hat{z}}}) \\ + \frac{1}{8} \sum_\mathbf{i} (D_x \mathbf{S_i} \times \mathbf{S_{i+2\hat{x}}} \cdot \hat{x} + D_y \mathbf{S_i} \times \mathbf{S_{i+2\hat{y}}} \cdot \hat{y} + D_z \mathbf{S_i} \times \mathbf{S_{i+2\hat{z}}} \cdot \hat{z}). \tag{7}$$

The simulation is performed on an $N = 24^3$ cubic lattice with period boundary conditions using the standard Metropolis algorithm [39] and the parallel tempering algorithm [40]. We take an exchange sampling after every 10 standard MC steps. Typically, the initial $6\times10^5$ steps are discarded for equilibrium consideration and another $6\times10^5$ steps are retained for statistic averaging of the simulation. Occasional checks were made on a larger lattice of up to 40 to ensure that finite-size effect never affect our conclusion. We analyze the spin structures by making the Fourier transform

$$\langle \mathbf{S_k} \rangle = \frac{1}{N} \sum_{\mathbf{i}} \langle \mathbf{S_i} \rangle e^{-i\mathbf{k}\cdot\mathbf{i}} \quad , \tag{8}$$

and calculating the intensity profile $|\langle \mathbf{S_k} \rangle|^2$. Furthermore, we also calculate the longitudinal susceptibility $\chi_z$, and the uniform chirality $\chi$

$$\chi = \frac{1}{8\pi} \sum_{\mathbf{i}} \left[ \mathbf{S_i} \cdot (\mathbf{S_{i+\hat{x}}} \times \mathbf{S_{i+\hat{y}}}) + \mathbf{S_i} \cdot (\mathbf{S_{i-\hat{x}}} \times \mathbf{S_{i-\hat{y}}}) \right] \tag{9}$$

to estimate the phase transition points [8].

**Results and discussion**

**Reorientation of the wave vector of the helix.** First, we study the case of the exchange and DM interaction anisotropies with the same magnitude at zero *h*. Generally, the anisotropy magnitude $\alpha$ and the ratio of the DM interaction to the exchange interaction $\beta$ are defined by:

$$\alpha = \frac{J_{x,y}}{J_z}, \beta = \frac{D_\mu}{J_\mu}. \tag{10}$$

Here, Eq. (4) is updated to:

$$E = -J_x \left[ (\cos k_x + \cos k_y + \frac{1}{\alpha} \cos k_z) + (\beta \sin\varphi(\cos\theta \sin k_x + \sin\theta \sin k_y) \right. \\ \left. + \frac{\beta}{\alpha} \cos\varphi \sin k_z) \right]. \tag{11}$$

Once the energy expression is minimized, we obtain (the modulus of **k** and energies *E*):

(1) the helical spin state with **k** = k(0, 0, 1)

$$k_z = \arctan(\beta) \text{ for } \varphi = 0, \tag{12}$$

$$E_{[001]} = -J_x(2 + \frac{\sqrt{1+\beta^2}}{\alpha}).$$

(2) the helical spin state with **k** = k(1, 1, 0)

$$k_{x,y} = \arctan(\frac{\beta}{\sqrt{2}}) \text{ for } \theta = \frac{\pi}{4}, \varphi = \frac{\pi}{2}, \tag{13}$$

$$E_{[110]} = -J_x(\sqrt{4+2\beta^2} + \frac{1}{\alpha}).$$

(3) the helical spin state with **k** = $(k_x, k_y, k_z)$

$$k_{x,y} = \arcsin\sqrt{\frac{\alpha^2(1+\beta^2)-1}{\alpha^2(\beta^2+3)}}, \quad k_z = \arcsin\sqrt{\frac{\beta^2 - 2\alpha^2 + 2}{\beta^2+3}}$$

$$\text{for } \varphi = \sqrt{\frac{2\alpha^2(1+\beta^2)-2}{\beta^2(2\alpha^2+1)}}, \quad \theta = \frac{\pi}{4}, \tag{14}$$

$$E_{[xxz]} = -J_x(\frac{3+\beta^2}{\alpha}\sqrt{\frac{2\alpha^2+1}{\beta^2+3}}).$$

Furthermore, the [xxz] helical state is limited by $0 < \varphi < \pi/2$ and

$$\sqrt{\frac{1}{1+\beta^2}} < \alpha < \sqrt{\frac{2+\beta^2}{2}}. \tag{15}$$

It is expected that $\alpha$ increases ($\alpha > 1$) when a compressive strain is applied along the [110] axis. Interestingly, the [110] helical order will win out over the [111] helical phase for $\alpha > \sqrt{1.5}$, as clearly shown in Fig. 1(a) which gives the calculated energies for a fixed $\beta = 1$. Thus, the stress-induced reorientation of the wave vector of the helix reported in experiments can be qualitatively reproduced in our anisotropic model. Similarly, the [111] helical order will be reproduced by the [001] one for small $\alpha < \sqrt{2}/2$, related to the case of compressive (tensile) stress applied along the [001] ([110]) axis, in some extent [41, 42]. The calculated ground-state phase diagram in the ($\alpha$, $\beta$) parameter plane is shown in Fig. 1(b) which can be divided into three parameter regions with different helical orders. It is noted that the helical propagation direction gradually moves towards the pressure axis ($\varphi$ gradually changes) when the magnitude of the anisotropy is increased, well consistent with experimental observation. Furthermore, the $\alpha$-region with the [xxz] helical order is extensively suppressed as $\beta$ decreases, demonstrating that the helical order in chiral magnet with a weak DM interaction can be easily modulated by the uniaxial stress [43].

As a matter of fact, these helical spin orders are also confirmed in our MC simulations. For example, the [001] helical order is stabilized at low $T$ for ($\alpha$, $\beta$) = (0.866, 0.577), and its spin configuration and the Bragg intensity are shown in Fig. 2(a). In one in-plane (xy) lattice layer, all the spins are parallel with each other. In addition, the spins of the chain along the [001] direction form a spiral structure, clearly demonstrating the helical order with the wave vector **k** = (0, 0, k). For $\alpha < 1$ (pressure applied along the [001] axis), the exchange interaction $J_z$ and DM interaction $D_z$ play an essential role in determining the ground-state of the system, and their competition results in the development of the [001] helical order. Thus, the compressive strain will tune the wave vector from the [111] axis to the stress axis, as reported in experiments. Similarly, the [110] helical order (Fig. 2(b)) and the [111] helical order (Fig. 2(c)) can be developed for ($\alpha$, $\beta$) = (1.155, 0.816) and ($\alpha$, $\beta$) = (1, 1), respectively. Furthermore, these spin orders can be also reflected in the calculated Bragg intensities, as given in the bottom of Fig. 2.

On the other hand, it is noted that the magnitude of the exchange anisotropy probably is not the same as that of the DM interaction anisotropy, especially when the spin-orbit coupling of the system is anisotropic [23]. Thus, this case is also investigated for integrity in this work. We define the following two parameters:

$$\gamma = \frac{D_{x,y}}{J_{x,y}}, \quad \xi = \frac{D_z}{J_z}. \tag{16}$$

Following the earlier work [23], a constraint $\frac{J_z}{J_{x,y}} = \sqrt{\frac{(1+\gamma^2)}{(1+\xi^2)}}$ is applied, and Eq. (4) is updated to:

$$E = -J_x \left[ (\cos k_x + \cos k_y + \sqrt{\frac{(1+\gamma^2)}{(1+\xi^2)}} \cos k_z) \right.$$
$$\left. - (\gamma \sin\varphi (\cos\theta \sin k_x + \sin\theta \sin k_y) + \sqrt{\frac{(1+\gamma^2)}{(1+\xi^2)}} \xi \cos\varphi \sin k_z) \right] . \quad (17)$$

Similarly, the phase boundaries can be exactly obtained by comparing these energies of the helical orders, and the calculated ground-state phase diagram in the ($\gamma$, $\gamma/\xi$) parameter plane is shown in Fig. 3. It is clearly demonstrated that these helical orders can be effectively modulated by these parameters, further strengthening the conclusion that the interaction anisotropy may be important in understanding the uniaxial pressure dependence of ground-state in chiral magnets such as MnSi.

**Variation of the *T*-region with the skyrmion lattice phase.** With a magnetic field applied along the [001] direction, the skyrmion lattice phase exists in a small *h-T* region. A transverse (longitudinal) pressure further stabilizes (destabilizes) the skyrmion lattice phase, resulting in the extension (suppression) of the *T*-region with this phase, as reported in earlier experiments on bulk MnSi [20, 21]. This behavior is also captured in this anisotropic spin model.

Fig. 4(a) shows the simulated phase diagram for ($\alpha$, $\beta$) = (1.155, 0.816). Even with the compensation term, the skyrmion lattice phase remains stable at low *T*, demonstrating the prominent role of the interaction anisotropies in modulating the skyrmion lattice phase. This phenomenon can be understood by analyzing the spin structures. For one spin chain along the *z* direction in the [110] helical order, all the spins are parallel with each other, and $J_z$ interaction is completely satisfied. Thus, there is no energy loss from the $J_z$ interaction due to the transition from the helical order to the tube skyrmion phase, resulting in the extension of the *T*-region with the skyrmion lattice phase. In some extent, this behavior is similar to that of the two-dimensional system in which a rather large *T*-region with the skyrmion lattice phase has been reported both experimentally [4, 5] and theoretically [44] as a result of the suppression of the competing conical phase. In Fig. 4(b), we show a snapshot (one in-plane lattice layer) of the skyrmion lattice phase and the Bragg intensity at $T = 0.07$ and $h = 0.46$. The skyrmion phase of hexagonal symmetry is clearly confirmed. It is noted that the magnitude of the anisotropy may be not so large in real materials, and the skyrmion lattice phase at $T \to 0$ predicted here has not been reported experimentally. However, this work indeed manifests the important role of the interaction anisotropy in modulating the skyrmions.

On the contrary, the extent of the skyrmion lattice phase is significantly suppressed for $\alpha < 1$, as shown in Fig. 5 which gives the phase diagram for $(\alpha, \beta) = (0.866, 0.5735)$. With the increase of $J_z$ ($\alpha$ decreases), the energy loss from the $J_z$ interaction due to the transition to the skyrmion lattice phase increases, resulting in the destabilization of the skyrmion lattice phase. As a matter of fact, earlier experiment revealed that an in-plane tensile strain destabilizes the skyrmion lattice phase [36], consistent with our simulations. Furthermore, it is clearly shown that the helical order is only stabilized at zero $h$, which can be explained analytically. The spins in an in-plane layer are parallel with each other in the [001] helical order, exhibiting a quasi-one-dimensional property. In this case, the energy of the conical phase under small $h$ can be written by

$$E_{\text{con}} = -D_z \sin k_z \sqrt{1 - \frac{2\cos^2 \phi}{1 + \cos k_z}} - J_z \cos k_z - h \cos \phi, \tag{18}$$

where $\phi$ is the cone half-angle (for the [001] helical order, $\phi = \pi/2$). Once the energy express is minimized, we obtain

$$h(1 + \cos k_z)\sqrt{1 - \frac{2\cos^2 \phi}{1 + \cos k_z}} = 2D_z \sin k_z \cos \phi. \tag{19}$$

Thus, it is clearly indicated that the [001] helical order can only be develop at zero $h$. In fact, earlier experiments have revealed that both the helical order and skyrmion lattice phase can be destabilized by the longitudinal pressure [20]. Here, our work suggests that the conical phase will completely replace the helical one under finite $h$ in the system with strong interaction anisotropies.

**Conclusion**

In conclusion, we have studied the uniaxial stress effects on the magnetic orders of bulk MnSi based on the spatially anisotropic spin model. Several experimental observations are qualitatively reproduced by the analytical calculation and Monte Carlo simulations of the model. It is suggested that the helical orders as well as the skyrmion lattice phase can be effectively modulated by the interaction anisotropy tuned by the applied pressure, especially for the system with a weak DM interaction. The present work may provide new insights into the understanding of the magnetic orders in the strained MnSi.

**Acknowledgements**:

This work is supported by the National Key Projects for Basic Research of China (Grant No. 2015CB921202), and the National Key Research Programme of China (Grant No. 2016YFA0300101), and the Natural Science Foundation of China (Grant No. 51332007), and the Science and Technology Planning Project of Guangdong Province (Grant No. 2015B090927006), and Special Funds for Cultivation of Guangdong College Students' Scientific and Technological Innovation (Grant No. pdjh2017b0138). X. Lu also thanks for the support from the project for Guangdong Province Universities and Colleges Pearl River Scholar Funded Scheme (2016).

**Author contributions**

J.C. and M.H.Q. conceived the research project and J.C. performed the computations. S.D. and J.M.L. discussed the physical mechanism and extension. X.B.L. and X.S.G. provided the technical support. All the authors discussed the results and commented on the manuscript. J.C. and M.H.Q. wrote the paper.

**Additional information**

**Competing financial interests**

The authors declare no competing financial interests.

**FIGURE CAPTIONS**

Fig.1. (a) The local energies as a function of $\alpha$. (b) The ground-state phase diagram in the space of $(\alpha, \beta)$.

Fig.2. A plot of the spin configurations projected on the *xy* plane (up) and projected on the *yz* plane (middle). At the bottom of each figure are the plots of the Bragg intensity from Fourier transition which shows the sets of helix vectors. The parameters are (a) $(\alpha, \beta) = (0.866, 0.577)$, (b) $(\alpha, \beta) = (1.155, 0.816)$, and (c) $(\alpha, \beta) = (1, 1)$ at $T = 0.01$.

Fig.3. The ground-state phase diagram in the space of $(\gamma, \gamma/\xi)$.

Fig.4. (a) The estimated phase diagram in the $(T, h)$ plane for $(\alpha, \beta) = (1.155, 0.816)$, and (b) A plot of the in-plane layer spin configuration for the tube skyrmion phase. The intensity profile is also given in the bottom of (b).

Fig.5. The estimated phase diagram in the $(T, h)$ plane for $(\alpha, \beta) = (0.866, 0.577)$.

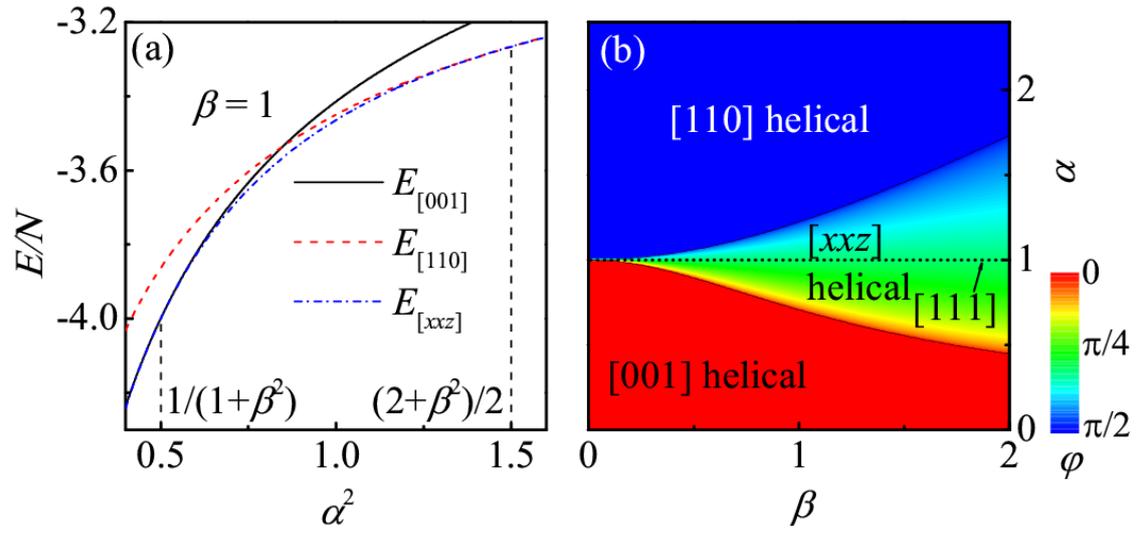

Fig.1. (a) The local energies as a function of $\alpha$. (b) The ground-state phase diagram in the space of ($\alpha$, $\beta$).

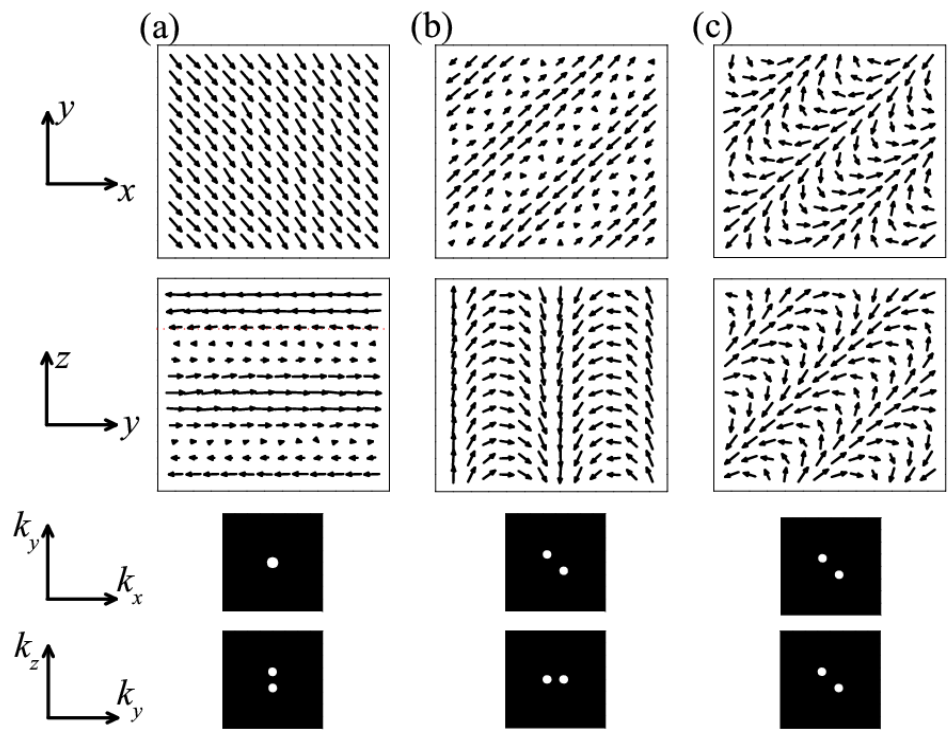

Fig.2. A plot of the spin configurations projected on the *xy* plane (up) and projected on the *yz* plane (middle). At the bottom of each figure are the plots of the Bragg intensity from Fourier transition which shows the sets of helix vectors. The parameters are (a) $(\alpha, \beta) = (0.866, 0.577)$, (b) $(\alpha, \beta) = (1.155, 0.816)$, and (c) $(\alpha, \beta) = (1, 1)$ at $T = 0.01$.

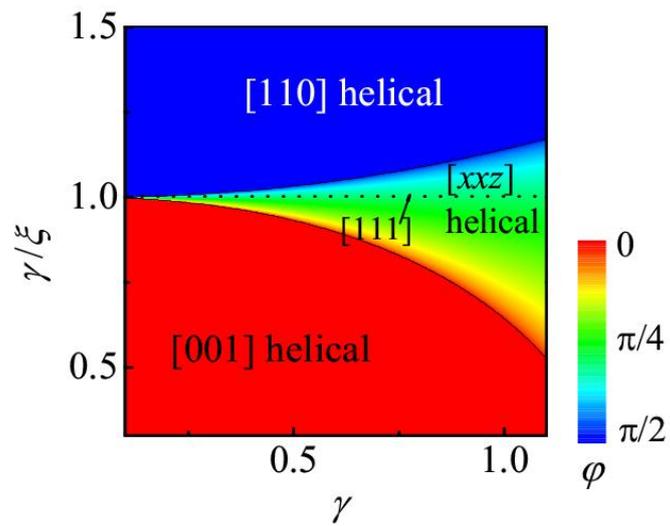

Fig.3. The ground-state phase diagram in the space of ($\gamma$, $\gamma/\xi$).

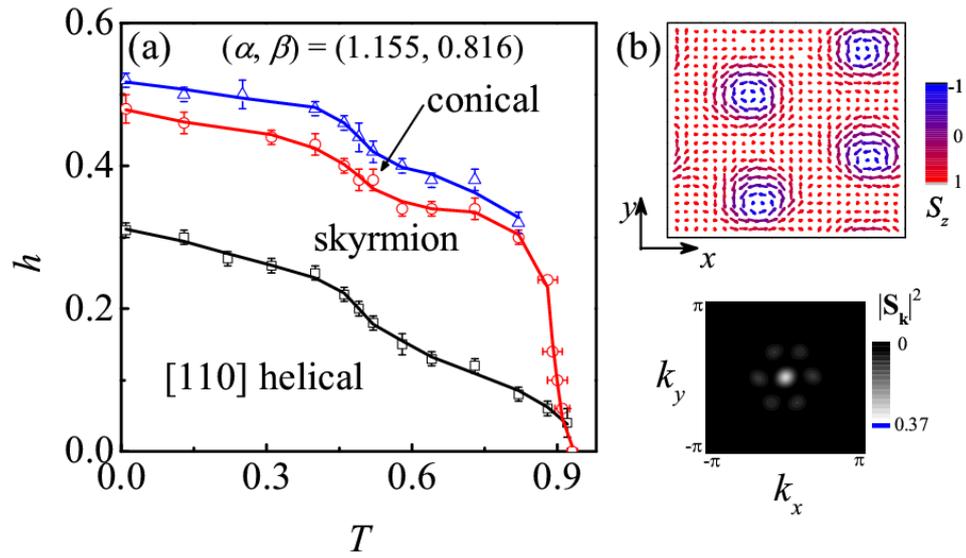

Fig.4. (a) The estimated phase diagram in the ($T$, $h$) plane for ($\alpha$, $\beta$) = (1.155, 0.816), and (b) A plot of the in-plane layer spin configuration for the tube skyrmion phase. The intensity profile is also given in the bottom of (b).

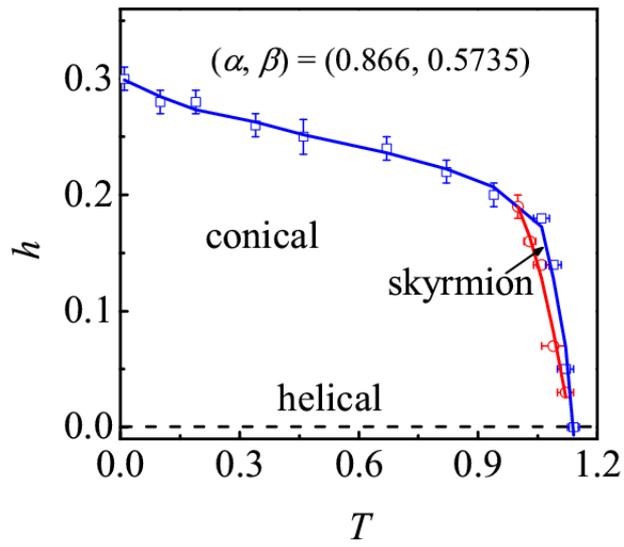

Fig.5. The estimated phase diagram in the (*T*, *h*) plane for ($\alpha$, $\beta$) = (0.866, 0.577).